\documentclass[aps,prb,twocolumn,showpacs,showkeys,floatfix]{revtex4}

\usepackage{graphicx,color}
\usepackage{multirow,slashbox}

\newcommand{\jwj}[1]{\textcolor{black}{#1}}
\graphicspath{{figs/}}
\begin{document}
\title{Graphene Versus MoS$_{2}$: a Mini Review}
\author{Jin-Wu Jiang}
    \altaffiliation{Email address: jiangjinwu@shu.edu.cn; jwjiang5918@hotmail.com}
    \affiliation{Shanghai Institute of Applied Mathematics and Mechanics, Shanghai Key Laboratory of Mechanics in Energy Engineering, Shanghai University, Shanghai 200072, People's Republic of China}

%\date{22 December 2009}
\date{\today}
\begin{abstract}

\jwj{Graphene and MoS$_{2}$ are two well-known quasi two-dimensional materials.} This review is a comparative survey of the complementary lattice dynamical and mechanical properties of graphene and MoS$_{2}$. This comparison facilitates the study of graphene/MoS$_{2}$ heterostructures, which is expected to mitigate the negative properties of each individual constituent.

\end{abstract}

\maketitle
\tableofcontents
%\pagebreak

\section{Introduction}
Quasi two-dimensional (Q2D) materials have many novel properties and have attracted intensive research interest over the past decades. The size of the family of Q2D materials keeps expanding. The Q2D family currently contains the following materials: graphene, hexagonal boron nitride, 2D honeycomb silicon, layered transition metal dichalcogenides (MoS$_{2}$, WS$_{2}$, ...), black phosphorus and 2D ZnO. Graphene is the most well-known material among the Q2D family of materials. Novoselov and Geim were awarded the Nobel prize in physics for graphene in 2010.\cite{GeimAK2007nm}

The investigations on graphene are extensive but not exhaustive.\cite{NetoAHC2011rpp} They are helpful for the whole Q2D family because many of the experimental set ups (initially for graphene) can be used to perform measurements on other materials in this family. For example, the mechanical properties of single-layer MoS$_{2}$ (SLMoS$_{2}$) were successfully measured using the same nanoindentation platform as graphene.\cite{LeeC2008sci,CooperRC2013prb1} In the theoretical community, many theorems or approaches, initially developed to study graphene, are also applicable to other Q2D materials. Some of the extensions may turn out to be trivial because of the common two-dimensional nature of these materials. However, the extensions may bring about new findings as these Q2D materials have different microscopic structures. For example, the bending modulus of SLMoS$_{2}$ can be derived using a similar analytic approach as graphene. The bending modulus of SLMoS$_{2}$ is about seven times larger than that of graphene, owing to its trilayer structure (one Mo layer sandwiched between two S layers).\cite{OuyangZC1997,TuZC2002,ArroyoM2002,LuQ2009,JiangJW2013bend} Another example is the puckered micro structure of black phosphorus, which leads to a negative Poisson's ratio in the out-of-plane direction.\cite{JiangJW2014bpnpr}

From the above, we found that graphene is attracting ongoing research interest from both the academic and applied communities. Many review articles have been devoted to graphene. \cite{GeimAK2007nm,FerrariAC2007ssc,NetoAHC2009rmp,GeimAK2009sci,MalardLM2009prrspl,RaoCNR,AllenMJ2010cr,BonaccorsoF2010npho,SchwierzF2010nn,Balandin2011nm} Addition, more and more researchers have begun examining possible applications of other Q2D materials, using the knowledge gained from graphene. In particular, MoS$_{2}$ has attracted considerable research interest. Many review articles have also been published on MoS$_{2}$.\cite{WangQH2012nn,ChhowallaM,Xu2013cr,Butler2013acsnn,HuangX2013csr}

The present review focusses on a detailed comparison of the mechanical properties of graphene and SLMoS$_{2}$. This comparison makes it clear as to what the positive/negative properties for graphene and MoS$_{2}$ are, highlighting the possible advanced features and drawbacks of graphene/MoS$_{2}$ heterostructures.\cite{BritnellL2013sci} These heterostructures were expected to mitigate the negative properties of each individual constituent. For example, graphene/MoS$_{2}$/graphene heterostructures have efficient photon absorption and electron-hole creation properties, because of the enhanced light-matter interactions in the SLMoS$_{2}$ layer.\cite{BritnellL2013sci} Another experiment recently showed that MoS$_{2}$ can be protected from radiation damage by coating it with graphene layers.\cite{ZanR2013acsn} This design exploits the outstanding mechanical properties of graphene.

In this review, we introduce and compare the following properties for graphene and the SLMoS$_{2}$; the structure, interatomic potential, phonon dispersion, Young's modulus, yield stress, bending modulus, buckling phenomenon, nanomechanical resonator, thermal conductivity, electronic band structure, optical absorption, and the graphene/MoS$_{2}$ heterostructure. This article ends up with a table that lists the major results for all of the properties that are compared in the article. 

\begin{figure}[tb]
  \begin{center}
    \scalebox{0.9}[0.9]{\includegraphics[width=8cm]{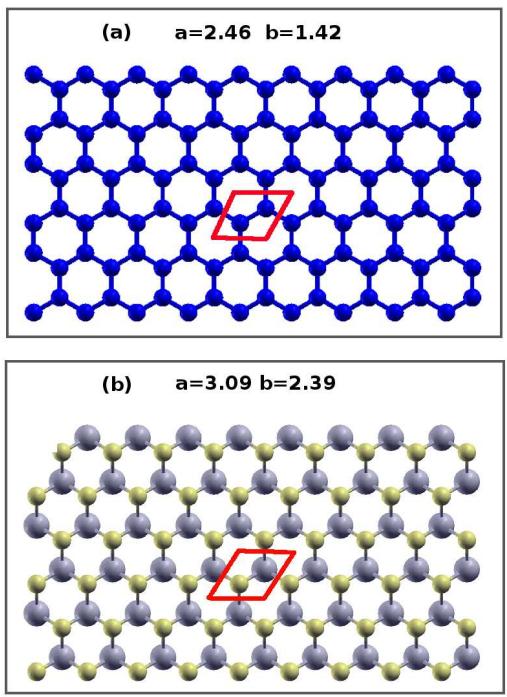}}
  \end{center}
  \caption{(Color online) Top view of the structures of (a) graphene and (b) SLMoS$_{2}$. The red rhombus encloses the unit cell in each structure. The numbers are the lattice constant ($a$) and the bond length ($b$) in \AA.}
 \label{fig_cfg}
\end{figure}

\section{Structure and Interatomic Potential}

\textbf{Structure.} First, the structures of graphene and SLMoS$_{2}$ are compared. Fig.~\ref{fig_cfg}~(a) shows that graphene has a honeycomb lattice structure with a $D_{6h}$ point group. There are two inequivalent carbon atoms in the unit cell. These two carbon atoms are reflected onto each other by the inverse symmetry operation from the $D_{6h}$ point group. The lattice constant is $a=2.46$~{\AA} and the C-C bond length is $b=a/\sqrt{3}=1.42$~{\AA}.\cite{SaitoR}

Figure ~\ref{fig_cfg}~(b) shows the top view of the SLMoS$_{2}$ structure. SLMoS$_{2}$ has a trilayer structure with one Mo atomic layer sandwiched between two outer S atomic layers. The small yellow balls represent the projection of the outer two S atomic layers onto the Mo atomic layer. The point group for SLMoS$_{2}$ is $D_{3h}$. The $\textbf{R}_\pi$ rotation symmetry is broken in SLMoS$_{2}$. There are two S atoms and one Mo atom in the unit cell. The lattice constant for the in-plane unit cell is $a=3.09$~{\AA} and the Mo-S bond length is $b=2.39$~{\AA}. These values were computed using the Stillinger-Weber (SW) potential.\cite{JiangJW2013sw} They agree with the first-principles calculations\cite{SanchezAM} and the experiments.\cite{WakabayashiN}

\textbf{Interatomic potential.} The interactions between the carbon atoms in graphene can be calculated using four different computation cost levels. The first principles calculation is the most expensive approach to compute the interatomic energy of graphene. Many existing simulation packages can be used for such calculations, including the commercial Vienna Ab-initio Simulation Package (VASP)\cite{KresseG1996prb} and the freely available Spanish Initiative for Electronic Simulations with Thousands of Atoms (SIESTA) package.\cite{siesta} To save on the computation cost, Brenner et al. developed an empirical potential for carbon-based materials, including graphene.\cite{brennerJPCM2002} The Brenner potential takes the form of the bond-order Tersoff potential\cite{TersoffJ4} and is able to capture most of the linear properties and many of the nonlinear properties of graphene. For instance, it can describe the formation and breakage of bonds in graphene, providing a good description of its structural, mechanical and thermal properties. The Tersoff potential\cite{TersoffJ4} or the SW potential\cite{StillingerF,AbrahamFF1989ss} provides reasonable predictions for some of the nonlinear and linear properties of graphene. These two empirical potentials have fewer parameters than the Brenner potential, thus they are faster than the Brenner potential. Last, the linear part of the C-C interactions in graphene can be captured using valence force field models (VFFMs).\cite{AizawaT} The VFFMs have the most inexpensive computation cost and can be used to efficiently compute some of the linear properties.

The potentials of these four computation levels can also be used for SLMoS$_{2}$. First principles calculations can also be used for SLMoS$_{2}$. In 2009, Liang et al. parameterized a bond-order potential for SLMoS$_{2}$,\cite{LiangT} which was based on the bond order concept underlying the Brenner potential.\cite{brennerJPCM2002} This Brenner-like potential was recently further modified to study the nanoindentations in SLMoS$_{2}$ thin films using a molecular statics approach.\cite{StewartJA} Recently, we parameterized the SW potential for SLMoS$_{2}$, where potential parameters were fitted to the phonon spectrum.\cite{JiangJW2013sw} This potential could easily be used in some of the popular simulation packages, such as the General Utility Lattice Program (GULP)\cite{gulp} and the Large-Scale Atomic/Molecular Massively Parallel Simulator (LAMMPS).\cite{lammps} In 1975, Wakabayashi et al.\cite{WakabayashiN} developed a VFFM to calculate the phonon spectrum in bulk MoS$_{2}$. This linear model has been used to study the lattice dynamical properties of some MoS$_{2}$ based materials.\cite{JimenezSS,DobardzicE,DamnjanovicM2008mmp}

\section{Phonon Dispersion}

\begin{figure}[tb]
  \begin{center}
    \scalebox{1}[1]{\includegraphics[width=8cm]{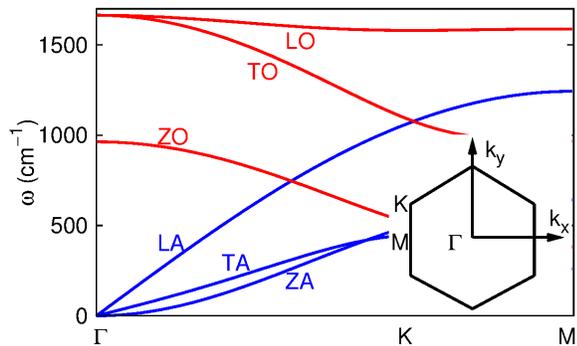}}
  \end{center}
  \caption{(Color online) Phonon dispersion of graphene along the high symmetry $\Gamma$KM lines in the Brillouin zone. The interactions between the carbon atoms were determined by the Brenner potential. The inset shows the first Brillouin zone for the hexagonal lattice structure.}
  \label{fig_phonon_dispersion_graphene}
\end{figure}

\begin{figure}[tb]
  \begin{center}
    \scalebox{1.0}[1.0]{\includegraphics[width=8cm]{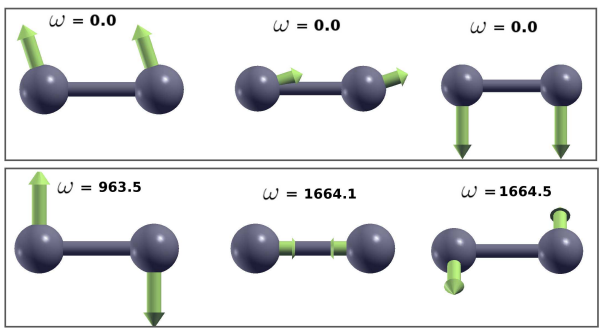}}
  \end{center}
  \caption{(Color online) Eigenvectors for the six phonon modes at the $\Gamma$ point in the first Brillouin zone of graphene. The arrow attached to each atom represents the vibration component of the atom in the eigenvector. The numbers are the frequencies of each phonon mode with units of cm$^{-1}$.}
  \label{fig_u_graphene}
\end{figure}

A phonon is a quasiparticle in reciprocal space. Each phonon mode describes a particular type of collective vibrations for all of the atoms in the real lattice space. 
The symmetry of the vibration morphology follows an irreducible representation of the space group of the system. These irreducible representations are denoted by the wave vector $\vec{k}$. The phonon modes are denoted by the wave vector $\vec{k}$ and the branch index $\tau$, where $\vec{k}$ is the inter-cell degree of freedom and $\tau$ corresponds to the intra-cell degree of freedom. Each phonon mode has a specific angular frequency ($\omega^{\tau}_{k}$) and eigenvector ($\vec{\xi}^{\tau}_{k}$). For graphene and SLMoS$_{2}$, each degree of freedom in the real lattice space can be indicated as $(l_{1} l_{2} s\alpha)$. $l_1$ and $l_2$ denote the position of the unit cell. $s$ describes the different atoms in the unit cell and $\alpha=x,y,z$ is the direction of the axis. The frequency and the eigenvector of the phonon mode can be obtained through the diagonalization of the following dynamical matrix,
\begin{eqnarray*}
&&D_{s\alpha;s'\beta}\left(\vec{k}\right) = \frac{1}{\sqrt{m_{s}m_{s'}}}\sum_{l_{1}=1}^{N_{1}}\sum_{l_{2}=1}^{N_{2}}K_{00s\alpha;l_{1}l_{2}s'\beta}e^{i\vec{k}\cdot\vec{R}_{l_{1}l_{2}}};\\
&&\\
&&\sum_{s'\beta}D_{s\alpha;s'\beta}\left(\vec{k}\right)\xi_{\beta}^{(\tau')}(\vec{k}|00s') = \omega^{(\tau)2}(\vec{k})\xi_{\alpha}^{(\tau')}(\vec{k}|00s).
\end{eqnarray*}
The force constant matrix $K_{00s\alpha;l_{1} l_{2} s'\beta}$ stores the information on the interactions between the two degrees of freedom, ($00s\alpha$) and ($l_{1} l_{2} s'\beta$). $N_1\times N_2$ gives the total number of unit cells. For the short-range interactions, a summation over $(l_1, l_2)$ can be truncated to the summation over the neighboring atoms.

\begin{figure}[tb]
  \begin{center}
    \scalebox{1}[1]{\includegraphics[width=8cm]{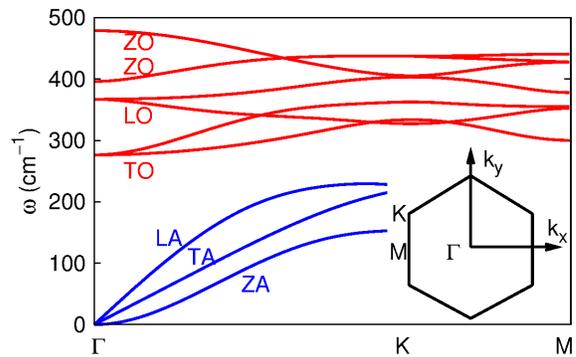}}
  \end{center}
  \caption{(Color online) Phonon dispersion of SLMoS$_{2}$ along the high symmetry $\Gamma$KM lines in the Brillouin zone. The interactions are described by the SW potential. The inset shows the first Brillouin zone for the hexagonal lattice structure.}
  \label{fig_phonon_dispersion_mos2}
\end{figure}

Figure ~\ref{fig_phonon_dispersion_graphene} shows the phonon dispersion of graphene along the high symmetry $\Gamma$KM lines in the first Brillouin zone. The force constant matrix was constructed using the Brenner potential.\cite{brennerJPCM2002} The inset shows the first Brillouin zone for the hexagonal lattice structure. There are six phonon branches in graphene according to the two inequivalent carbon atoms in the unit cell. These branches (from bottom to top) are the z-directional acoustic (ZA), transverse acoustic (TA), longitudinal acoustic (LA), z-directional optical (ZO), transverse optical (TO), and longitudinal optical (LO) branches. The three blue curves in the lower frequency range correspond to the three acoustic branches, while the upper three red curves correspond to the optical branches. The eigenvectors of the six phonon modes at the $\Gamma$ point in the first Brillouin zone of graphene are displayed in Fig.~\ref{fig_u_graphene}. In the top panel, the three acoustic phonon modes have zero frequency, indicating that the interatomic potential did not vary during the rigid translational motion. In the bottom panel, the two in-plane optical phonon modes have almost the same frequency, revealing the isotropic phonon properties for the two in-plane directions in graphene.\cite{WangH2009jrs}

\begin{figure}[tb]
  \begin{center}
    \scalebox{1.1}[1.1]{\includegraphics[width=8cm]{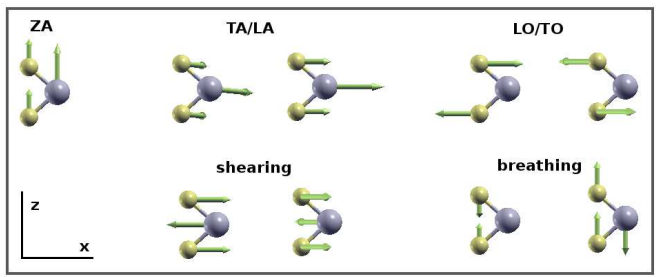}}
  \end{center}
  \caption{(Color online) Eigenvectors for the nine phonon modes at the $\Gamma$ point in the first Brillouin zone of SLMoS$_{2}$. There are three acoustic phonon modes, two intra-layer optical modes, two intra-layer shearing modes and two intra-layer breathing modes. The arrows attached to each atom represent the vibration component of the atom in the eigenvector.}
  \label{fig_u_mos2}
\end{figure}

Figure ~\ref{fig_phonon_dispersion_mos2} shows the phonon dispersion of SLMoS$_{2}$ along the high symmetry $\Gamma$KM lines in the first Brillouin zone. The atomic interactions are described by the SW potential.\cite{JiangJW2013sw} The inset shows the same first Brillouin zone as graphene. Each unit cell has one Mo atom and two S atoms, thus there are nine branches in the phonon spectrum. The three lower blue curves correspond to the three acoustic branches, while the six upper curves correspond to the optical branches. Fig.~\ref{fig_u_mos2} shows the eigenvectors for the nine phonons at the $\Gamma$ point in the first Brillouin zone of SLMoS$_{2}$. There are two interesting shear-like phonon modes and two inter-layer breathing-like phonon modes, as shown in the second raw.\cite{ZhangXprb2013}

From the phonon dispersion of graphene and SLMoS$_{2}$, it is difficult to determine which material has better phonon properties. Yet, there are two obvious differences in their phonon dispersions. First, the spectrum of graphene is overall higher than that of SLMoS$_{2}$ by about a factor of three. As a result, the phonon modes in graphene can carry more energy than those in SLMoS$_{2}$ in the thermal transport phenomenon, leading to the stronger thermal transport ability of graphene. Second, there is a distinct energy band gap between the acoustic and optical branches in SLMoS$_{2}$. This band gap forbids many phonon-phonon scattering channels; thus it protects the acoustic phonon modes from being interrupted by the high-frequency optical phonons in SLMoS$_{2}$.\cite{JiangJW2013mos2resonator} As a result, SLMoS$_{2}$ nanoresonators have a higher quality (Q)-factor than graphene, since the resonant oscillations in SLMoS$_{2}$ (related to the ZA mode) are affected by weaker thermal vibrations.

\section{Mechanical Properties}
The mechanical properties for both graphene and SLMoS$_{2}$ have been extensively investigated.\cite{huangPRB2006,LiuF2007prb,NiZ2009pbcm,GaoY2009peldsn,GuoYF2010prb,ShenL2010md,HaoF2011apl,ZhengY2011nano,HanT2011am,WeiY2012nm,ZhangYP2012drm,XuL2012jmc,YueQ2012pla} Here, several of the basic mechanical properties, including the Young's modulus, yield stress, bending modulus and buckling phenomenon will be discussed. These mechanical properties are fundamental for the application of graphene or SLMoS$_{2}$ in nano-devices. A good mechanical stability is essential in nanoscale devices, as they are sensitive to external perturbations because of their high surface to volume ratio.

\textbf{Young's modulus.} Here, the effective Young's modulus ($E^{2D}$), which is thickness independent, is discussed. The Young's modulus is related to this quantity through $Y=E^{2D}/h$, where $h$ is the film thickness. The thicknesses were chosen to be 3.35~{\AA} and 6.09~{\AA} for graphene and SLMoS$_{2}$, respectively. The nanoindentation experiments measured the effective Young's modulus of graphene to be around 335.0~{Nm$^{-1}$}.\cite{LeeC2008sci} This value could be reproduced using a simple approach, in which the nonlinear interactions are estimated from the Tersoff-Brenner potential.\cite{JiangJW2010young}

For SLMoS$_{2}$, similar nanoindentation experiments obtained an average value for the effective Young's modulus of $180\pm60$~{Nm$^{-1}$} in the experiment by Bertolazzi et al.,\cite{BertolazziS} and $120\pm 30$~{Nm$^{-1}$}, measured by Cooper et al.\cite{CooperRC2013prb1,CooperRC2013prb2} Recently, Liu et al. performed similar nanoindentation experiments on chemical vapor deposited SLMoS$_{2}$, obtaining an effective Young's modulus of about 170~{Nm$^{-1}$}.\cite{LiuK2014arxiv} The nanoindentation set up has also been used to study the Young's modulus of thicker MoS$_{2}$ films.\cite{gomezAM2012} The theoretical prediction of the effective Young's modulus is 139.5~{Nm$^{-1}$} for SLMoS$_{2}$, based on the SW potential.\cite{JiangJW2013sw}

\textbf{Yield stress.} The nanoindentation measurements can also determine the yield stress ($\sigma_{\rm int}$, the maximum of the stress-strain curve). Lee et al. determined the yield stress to be $42\pm 4$~{Nm$^{-1}$} for graphene.\cite{LeeC2008sci} The yield stresses obtained with the continuum elasticity theory were 42.4~{Nm$^{-1}$} using a tight-binding atomistic model,\cite{CadelanoE2009prl} and 44.4~{Nm$^{-1}$} using the Brenner potential.\cite{ReddyCD2006nano} While the elasticity continuum simulation gave an isotropic value for the yield stress in graphene, microscopic atomic models have predicted the yield stress to be chirality dependent in graphene. The first-principles calculations predicted the yield stress to be 40.5~{Nm$^{-1}$} in the zigzag direction and 36.9~{Nm$^{-1}$} in the armchair direction in graphene.\cite{LiuF2007prb} Molecular mechanics simulations obtained a yield stress of 36.9~{Nm$^{-1}$} in the zigzag direction and 30.2~{Nm$^{-1}$} in the armchair direction in graphene.\cite{ZhaoH2009nl} Both of the atomic models showed that graphene has a higher yield stress in the zigzag direction than in the armchair direction. Note that the definition of the armchair/zigzag direction is the opposite in Refs.~\onlinecite{LiuF2007prb} and ~\onlinecite{ZhaoH2009nl}. We have retained the definition from Ref.~\onlinecite{ZhaoH2009nl}, where the armchair direction is along the direction of the carbon-carbon bonds.

In SLMoS$_{2}$, nanoindentation experiments found that the yield stress was $15\pm 3$~{Nm$^{-1}$}, determined by Bertolazzi et al.,\cite{BertolazziS} and $16.5\pm$~{Nm$^{-1}$}, determined by Cooper et al.\cite{CooperRC2013prb1,CooperRC2013prb2} In the first-principles calculations, the yield stress was predicted to be 17.5~{Nm$^{-1}$} under a biaxial strain in SLMoS$_{2}$.\cite{TaoP2014jap} While the work on the yield stress in SLMoS$_{2}$ is limited at the present stage, considerable attention has been paid to the novel structure transition in SLMoS$_{2}$.\cite{LinYC2013,JiangJW2014mos2bandgap,KanM2014jpcc,DangKQ2014sm} In this structure transition, the outer two S atomic layers are shifted relative to each other, leading to abrupt changes in the electronic and phonon properties in SLMoS$_{2}$. This structure transition is the result of the trilayer configuration of SLMoS$_{2}$ and is not observed in graphene.

\textbf{Bending modulus.} Graphene is extremely soft in the out-of-plane direction, owing to its one-atom-thick structure.\cite{xinPRB2000,MaT2011apl,ShenY2012apl,ShiX2012apl,WeiY2013nl} Graphene is so thin that it has an extremely small bending modulus, which can be explained by the well-known relationship in the shell theorem, $D=E^{2D}h^2/(12(1-\nu^2))$, where $h$ is the thickness and $\nu$ is Poisson's ratio. The bending modulus of graphene has been derived analytically from two equivalent approaches; it was 1.17~{eV} using the geometric approach with the interactions described by the VFFM,\cite{OuyangZC1997,TuZC2002} and 1.4~{eV} from the exponential Cauchy-Born rule using the Brenner potential.\cite{ArroyoM2002,LuQ2009} Note that these two approaches are equivalent to each other and the difference in the bending modulus mainly comes from the different potentials used in these two studies.

A similar analytic approach was used to derive the bending modulus of SLMoS$_{2}$ using the SW potential.\cite{JiangJW2013bend} The bending modulus of SLMoS$_{2}$ is 9.61~{eV}, which is larger than that of graphene by a factor of seven. The large bending modulus of SLMoS$_{2}$ is due to its trilayer atomic structure, which results in more interaction terms inhibiting the bending motion. The bending modulus can be calculated using:
\begin{eqnarray}
D & = & \frac{\partial^{2}W}{\partial \kappa ^{2}},
\label{eq_D}
\end{eqnarray}
where $W$ is the bending energy density and $\kappa$ is the bending curvature. For SLMoS$_{2}$, the bending energy can be written as:\cite{JiangJW2013bend}
\begin{eqnarray}
D & = & \sum_{q}\frac{\partial^{2}W}{\partial r_{q}^{2}}\left(\frac{\partial r_{q}}{\partial\kappa}\right)^{2}+\sum_{q}\frac{\partial^{2}W}{\partial\theta_{q}^{2}}\left(\frac{\partial\theta_{q}}{\partial\kappa}\right)^{2},
\label{eq_D2}
\end{eqnarray}
where $r_q$ and $\theta_q$ are the geometrical parameters in the empirical potential expressions. This formula is substantially different from the bending modulus formula for graphene.\cite{ArroyoM2004} Specifically, the first derivatives, $\frac{\partial r_{q}}{\partial\kappa}$ and $\frac{\partial\theta_{q}}{\partial\kappa}$, are nonzero, owing to the trilayer structure of SLMoS$_{2}$. As a result, the bending motion in SLMoS$_{2}$ will counteract with an increasing amount of cross-plane interactions.

\textbf{Buckling phenomenon.} The buckling phenomenon can be a disaster in electronic devices, but it can also be useful in some situations.\cite{ZhangJ2010prl,WangY2011acsnn,ShiJX2011cms,WangC2013pccp} The Euler buckling theorem states that the buckling critical strain can be determined from the effective Young's modulus and the bending modulus through following formula:\cite{TimoshenkoS1987}
\begin{eqnarray}
\epsilon_{c}  =  -\frac{4\pi^{2}D}{E^{2D}L^{2}},
\label{eq_euler}
\end{eqnarray}
where $L$ is the length of the system. For graphene, $E^{2D}=335$~{Nm$^{-1}$} and $D=1.4$~{eV}, found from the above discussions, giving the explicit formula for the buckling critical strain:
\begin{eqnarray}
\epsilon_{c}  =  -\frac{2.64}{L^{2}}.
\label{eq_euler_graphene}
\end{eqnarray}
The length ($L$) has units of \AA.

For SLMoS$_{2}$, $E^{2D}=139.5$~{Nm$^{-1}$} and $D=9.61$~{eV} were determined from the above discussions. The explicit formula for the buckling critical strain can is:
\begin{eqnarray}
\epsilon_{c}  =  -\frac{43.52}{L^{2}}.
\label{eq_euler_mos2}
\end{eqnarray}
The buckling critical strain for SLMoS$_{2}$ is twenty times larger than graphene, for samples of the same length; i.e., it is difficult to buckle SLMoS$_{2}$ under external compression. This phenomenon has been examined with both MD simulations and phonon analysis.\cite{JiangJW2014mos2bandgap,JiangJW2014mos2buckling}

We have discussed and compared the mechanical properties of graphene and SLMoS$_{2}$. Graphene has a larger Young's modulus and a larger yield stress and is more flexible than SLMoS$_{2}$. However, SLMoS$_{2}$ has a higher bending modulus and does not buckle as readily under external compression as graphene. Hence, in terms of the mechanical properties, it is more advantageous for graphene and SLMoS$_{2}$ to be used together in a heterostructure, to mitigate the negative mechanical properties of each constitute.

\section{Nanomechanical Resonators}
Nanoresonators based on two-dimensional materials such as graphene and SLMoS$_{2}$ are promising candidates for ultra-sensitive mass sensing and detection because of their large surface areas and small masses.\cite{HeXQ2005nano,ZhouM2009ac,XuY2010apl,LiuY2011jmps,WangJ2011jpdap,XuY2011nrl,GuF2011aps,ShenZB2012cms,ZhouSM2014cms} For sensing applications, it is important that the nanoresonator exhibits a high Q-factor because the sensitivity of a nanoresonator is inversely proportional to its Q-factor.\cite{EkinciKL} The Q-factor is a quantity that records the total number of oscillation cycles of the resonator before its resonant oscillation decays considerably. Hence, a weaker energy dissipation leads to a higher Q-factor.

For graphene nanoresonators, the Q-factor increases exponentially with a decreasing temperature\cite{ZandeAMVD,ChenC2009nn}, $T^{-\alpha}$. Zande et al.\cite{ZandeAMVD} found that the exponent $\alpha$=$0.35\pm0.05$ for temperatures below 40~K. For temperatures above 40~K, $\alpha=2.3\pm 0.1$. Chen et al.\cite{ChenC2009nn} observed a similar transition in the Q-factor. This continuous transition for the temperature dependence of the Q-factor is attributed to the diffusion of adsorbs in the out-of-plane direction on the surface of the graphene layer.\cite{JiangJW2013diffusion,EdblomC2014arxiv} The MD simulations also predicted a discontinuous transition in the Q-factor at the low temperature of 7.0~K, caused by the in-plane diffusion of adsorbs on the graphene surface.\cite{JiangJW2013diffusion} A very high Q-factor has been achieved in the laboratory at low temperatures. Bunch \emph{et al.} observed a Q-factor of 9000 for a graphene nanoresonator at 10~K.~\cite{ZandeAMVD}. Chen {\it et al.} also found that the Q-factor increased with decreasing temperature, reaching $10^{4}$ at 5~K~\cite{ChenC2009nn}. Eichler {\it et al.}~\cite{EichlerA} found that the Q-factor of a graphene nanoresonator reached $10^{5}$ at 90~mK.

For SLMoS$_{2}$, two recent experiments demonstrated the nanomechanical resonant behavior in SLMoS$_{2}$\cite{Castellanos-GomezA2013adm} and few-layer MoS$_{2}$\cite{LeeJ2013acsnano}. Castellanos-Gomez et al. found that the figure of merit, i.e., the frequency-Q-factor product is $f_{0}\times Q\approx 2\times10^{9}$ Hz for SLMoS$_{2}$.\cite{Castellanos-GomezA2013adm} Lee et al. found that few-layer MoS$_{2}$ resonators exhibit a high figure of merit of $f_{0}\times Q\approx 2\times10^{10}$ Hz.\cite{LeeJ2013acsnano} The high Q-factor of SLMoS$_{2}$ was attributed to the energy band gap in the phonon dispersion of SLMoS$_{2}$, which protects the resonant oscillations from being scattered by thermal vibrations.\cite{JiangJW2013mos2resonator} As a result, it was predicted that the Q-factor of SLMoS$_{2}$ nanoresonators will be higher than the graphene nanoresonators by at least a factor of four.

Although it has been theoretically predicted that MoS$_{2}$ will have better mechanical resonance behavior than graphene, experiments on MoS$_{2}$ nanoresonators are limited. More measurements are needed to examine their properties, such as the mass sensitivity. \jwj{Furthermore, the sensor application of the nanoresonators depends on the level of low frequency 1/f noise, which is a limiting factor for the communication applications and the sensor sensitivity and selectivity for the graphene and MoS$_{2}$ nanoresonators.\cite{LinYM2008nl,PalAN2009prl,ChengZ2010nl,RumyantsevS2010jpcm,LiuG2012apl,HossainMZ2013apl,BalandinAA2013nn}}

\section{Thermal Conductivity}
The thermal transport phenomenon occurs when there is a temperature gradient in a material. The thermal energy can be carried by phonons or electrons. The electronic thermal conductivity is important for metals. For graphene, the thermal conductivity is mainly contributed to by phonons, while the electronic thermal conductivity is less than 1\% of the overall thermal conductivity in graphene.\cite{SaitoK,YigenS} Thus, only the phonon (lattice) thermal conductivity will be discussed for  the graphene. \jwj{The thermal transport is in the ballistic regime at low temperature with weak phonon-phonon scattering.\cite{WangJSnegf,WangJS2013fp} For ballistic transport, the thermal conductivity ($\kappa$) is proportional to the length ($L$) of the system, $\kappa\propto L$. At high temperature, the phonon-phonon scattering is strong, so the thermal transport is in the diffusive regime. For diffusive transport, the thermal conductivity is related to the thermal current density ($J$) and the temperature gradient ($\bigtriangledown T$) through the Fourier law, $\kappa=-\bigtriangledown T/J$.}

\jwj{In bulk materials, the room temperature thermal conductivity is normally a constant that is size independent. Graphene has high thermal conductivity,\cite{ChenSS,GuoZ2009apl,XuY2009apl,HaoF2011apl,ChenS2011acsnn,WeiZ2011carbon,XieZ2011jpcm,ZhaiX2011epl,PenXF2011apl,MaF2012apl,WeiN,GuoZX} which behaves irregularity with the length in Q2D graphene; i.e., the in-plane thermal conductivity is not constant and increases with an increasing sample length.\cite{MingoN2005prl,MingoN2005nl,NikaDL2009prb,NikaDL2011pssb,WangJ2012jpcm,NikaDL2012jpcm,NikaDL2012nl,LiNB,XuX2014nc} For 10~nm long graphene, the room temperature thermal conductivity from the MD simulation is on the order of\cite{JiangJW2010isotopic} 60~{Wm$^{-1}$K$^{-1}$}. It increased quickly with an increasing length and reached 250~{Wm$^{-1}$K$^{-1}$} for 300~nm long graphene.\cite{XuX2014nc} For a length of 4.0~{$\mu$m}, graphene had a thermal conductivity around\cite{CaiWW2010nl} 2500~{Wm$^{-1}$K$^{-1}$}. For graphene samples larger than 10~{$\mu$m}, the thermal conductivity varies in the range from 1500~{Wm$^{-1}$K$^{-1}$} to 5000~{Wm$^{-1}$K$^{-1}$} depending on the sample size and quality.\cite{BalandinAA2008,GhoshS2008apl,XuX2014nc} The highest value of 5000~{Wm$^{-1}$K$^{-1}$} was achieved for the 20~{$\mu$m} sample by Balandin et al. These studies show that the thermal conductivity in graphene increases with an increasing dimension, although the sample size is larger than the phonon mean free path.\cite{GhoshS2008apl} However, there is still no universally accepted fact on the underlying mechanism for the size dependence of the thermal conductivity in graphene.\cite{LindsayL,NikaDL2009prb,AksamijaZ,NikaDL2012jpcm,ChenL2012jap,WeiZ2014jap} In the out-of-plane direction, it has been shown that the thermal conductivity for graphene decreases with increasing layer number, as there are more phonon-phonon scattering channels in thicker few-layer graphene.\cite{GhoshS,LindsayL,SinghD,ZhangG2011nns,ZhongWR20111,ZhongWR20112,RajabpourA,CaoHY,SunT}}

For MoS$_{2}$, a recent experiment by Sahoo et al. found that few-layer MoS$_{2}$ has a thermal conductivity around\cite{SahooS} 52~{Wm$^{-1}$K$^{-1}$}, which is much lower than that in thick graphene layers (1000~{Wm$^{-1}$K$^{-1}$}).\cite{GhoshS} Although there are currently no measurements on the thermal conductivity of SLMoS$_{2}$, this topic has attracted increasing interest in the theoretical community.\cite{VarshneyV,HuangW,JiangJW2013mos2} In 2010, Varshney et al. performed force-field based MD simulations to study the thermal transport in SLMoS$_{2}$.\cite{VarshneyV} In 2013, two first-principles calculations were performed to investigate the thermal transport in SLMoS$_{2}$ in the ballistic transport regime.\cite{HuangW,JiangJW2013mos2} The predicted room temperature thermal conductivity in the ballistic regime was below 800~{Wm$^{-1}$K$^{-1}$} for a 1.0~{$\mu$m} long SLMoS$_{2}$ sample.\cite{JiangJW2013mos2} This value is considerably lower than the ballistic thermal conductivity of 5000~{Wm$^{-1}$K$^{-1}$} for a graphene sample with the same length.\cite{JiangJW2009direction} The smaller thermal conductivity of SLMo$_{2}$ in the ballistic regime was caused by the lower phonon spectrum in SLMoS$_{2}$, which is lower than that of graphene by a factor of three. Each phonon mode in SLMoS$_{2}$ carries less thermal energy than in graphene. In 2013, we performed MD simulations to predict the room temperature thermal conductivity of SLMoS$_{2}$, which was 6.0~{Wm$^{-1}$K$^{-1}$} for a 4.0~{nm}-long system.\cite{JiangJW2013sw} More recently, the size dependence of the thermal conductivity in SLMoS$_{2}$ was studied with MD simulations. The value obtained was below 2.0~{Wm$^{-1}$K$^{-1}$} for a system with a length shorter than 120.0~nm.\cite{LiuX2013apl}

The above discussions have established that graphene has a much higher thermal conductivity than SLMoS$_{2}$. \jwj{The high thermal conductivity of graphene is useful for transporting heat out of the electronic transistor devices, so graphene can be used to enhance the thermal transport capability of some composites.\cite{YangZ,GoyalV,ShahilKMF,YuW2010nano,YuW2010jap,SongP2011polymer,YuaW2011pla,GoliP2014jps,MalekpourH2014nl}} Current transistors operate at very high speeds and are damaged by the inevitable Joule heating if the generated heat energy is not pumped out effectively. In this sense, graphene has better thermal conductivity properties than SLMoS$_{2}$.

\section{Electronic Band structure}
The electronic band structure is fundamental for electronic processes, such as the transistor performance. In particular, the value of the electronic band gap determines whether the material is metallic (zero band gap), semiconductor (moderate band gap), or insulator (large band gap).

Graphene has a nice electronic properties.\cite{WangY2009jpcc} Electrons in graphene have a linear energy dispersion near the Brillouin zone corner, which are massless Dirac fermions with 1/300 the speed of light.\cite{NovoselovKS2005nat,ZhouSY2006np} The Dirac fermion was found to be closely related to the mirror plane symmetry in the AB-stacked few-layer graphene; i.e. Dirac fermions present in AB-stacked few-layer graphene with an odd number of layers and the electronic spectrum becomes parabolic in AB-stacked few-layer graphene with an even number of layers.\cite{PartoensB2007prb} Interestingly, Dirac fermions present again in twisted bilayer graphene.\cite{HassJ2008prl} It is due to the effective decoupling of the two graphene layers by the twisting defect; i.e., the mirror plane symmetry is effectively recovered in the twisted bilayer graphene. The Dirac cone at the Brillouin zone corner has a zero band gap in graphene, that is mainly contributed by free $\pi$ electrons.\cite{ReichS2002prb} For electronic device, like transistors, a finite band gap is desirable, so various techniques have been invented to open an electronic band gap in graphene. The strain engineering can generate finite band gap of 0.1~{eV} for a 24\% uniaxial strain.\cite{PereiraVM2009prb} Guinea et al. applied tirangular symmetric strains to generate a band gap over 0.1~{eV}, which is observable at room temperature.\cite{GuineaF2010natp} A finite band gap can also be opened by confining the graphene structure in a nanoribbon form, where the band gap increases with decreasing ribbon width.\cite{NakadaK1996prb}

Electrons in SLMoS$_{2}$ are normal fermions with parabolic energy dispersion, and it is a semiconductor with a direct band gap above 1.8~{eV}.\cite{LiY2008jacs,KamKK,LuP2012pccp,EknapakulT2014nl} This finite band gap endorses SLMoS$_{2}$ to work as a transistor.\cite{RadisavljevicB2011nn,SangwanVK} Similar as graphene, the band gap in SLMoS$_{2}$ can also be modulated by strain engineering. First principles calculations predict a semiconductor-to-metal transition in SLMoS$_{2}$ by both biaxial compression or tension.\cite{EmilioS2012nnr} The experiment by Eknapakul et al. shows that an uniaxial tensile mechanical strain of 1.5\% can produce a direct-to-indirect band gap transition.\cite{ConleyHJ} With increasing number of layers, the electronic band gap for few-layer MoS$_{2}$ undergoes a direct-to-indirect transition, and decreases to a value of 1.2~{eV} for bulk MoS$_{2}$.\cite{MakKF}

From these comparisons, we find that SLMoS$_{2}$ possesses a finite band gap prior to any gap-opening engineering. Consequently, it may be more competitive than graphene for  applications in transistor, optoelectronics, energy harvesting, and other nano-material fields.

\section{Optical Absorption}
Optical properties of Q2D materials are important for their applications in photodetector, phototransitor, or other photonic nanodevices. The photocarriers in these Q2D materials may have quite different behavior from conventional semiconductors, due to their particular configuration.

Graphene has a Dirac cone electron band structure with zero band gap.\cite{NovoselovKS2005nat,ZhouSY2006np} Relating to this unique band structure, graphene can absorb about 2\% of incident light over a broad wavelength, which is strong considering its one-atom-thick nature.\cite{NairRR2008sci} Xia et al. demonstrated an ultra fast photodetector behavior for graphene, where the photoresponse did not degrade for optical intensity modulations up to 40~{GHz}, and the intrinsic band width was estimated to be above 500~{GHz}.\cite{XiaF2009nn} However, the photoresponsivity for graphene is low due to zero bandgap.

SLMoS$_{2}$ has a direct band gap about 1.8~{eV}.\cite{KamKK,EknapakulT2014nl} This optical-range band gap lead to high absorption coefficient for incident light, so the SLMoS$_{2}$ have very high sensitivity in photon detection.\cite{MakKF} Lopez-Sanchez et al. found that the photoresponsivity of the SLMoS$_{2}$ can be as high as 880~{AW$^{-1}$} for an incident light at the wavelength of 561~nm, and the photoresponse is in the 400-680~nm.\cite{Lopez-SanchezO2013nn} This high photoresponsivity together with its fast light emission enables SLMoS$_{2}$ to be an ultra sensitive phototransistors with good device mobility and large ON current. In phototransistors, the electron-hole pair can be efficiently generated by photoexcition in doped SLMoS$_{2}$, which joins the doping-induced charges to form a bound states of two electrons and one hole. As a result, the carrier effective mass is considerably increased, and the photoconductivity can be decreased.\cite{LuiCH2014}

For optical properties, graphene is very fast in the photo detection, while SLMoS$_{2}$ is very sensitive in this photo detection application. Considering this complementary property, it may be fruitful for the cooperation of these two materials.

\begin{table*}
\caption{The properties that have been compared for graphene and SLMoS$_{2}$ in the present review article.}
\label{tab_summary}

\begin{tabular}{|l|c|c|}
\hline 
properties & graphene & SLMoS$_{2}$\tabularnewline
\hline 
structrue & $D_{6h}$; $a=2.46\AA$; $b=1.42\AA$ (Ref.~\onlinecite{SaitoR}) & $D_{3h}$; $a=3.09\AA$; $b=2.39\AA$ (Ref.~\onlinecite{JiangJW2013sw})\tabularnewline
\hline 
interaction & $ab$ initio; Brenner; SW; VFFM & $ab$ initio; Brenner; SW; VFFM\tabularnewline
\hline 
phonon dispersion & $\omega_{\rm op}\approx1664.5$ cm$^{-1}$; $\omega_{\rm gap}=0$ & $\omega_{\rm op}\approx478.8$ cm$^{-1}$; $\omega_{\rm gap}=25.0$ cm$^{-1}$\tabularnewline
\hline 
\multirow{4}{*}{Young's modulus} & & $E^{2D}=180\pm60$ Nm$^{-1}$ (Ref.~\onlinecite{BertolazziS}) \tabularnewline
 & $E^{2D}=335.0$ Nm$^{-1}$ (Ref.~\onlinecite{LeeC2008sci}) & $E^{2D}=120\pm30$ Nm$^{-1}$ (Ref.~\onlinecite{CooperRC2013prb1}) \tabularnewline
 &  & $E^{2D}=170$ Nm$^{-1}$  (Ref.~\onlinecite{LiuK2014arxiv})\tabularnewline
 &  & $E^{2D}=139.5$ Nm$^{-1}$ (Ref.~\onlinecite{JiangJW2013sw}) \tabularnewline
\hline 
\multirow{5}{*}{yield stress} & $\sigma_{\rm int}=42\pm4$ Nm$^{-1}$ (Ref.~\onlinecite{LeeC2008sci}) & $\sigma_{\rm int}=15\pm3$ Nm$^{-1}$ (Ref.~\onlinecite{BertolazziS}) \tabularnewline
 & $\sigma_{\rm int}=42.4$ Nm$^{-1}$  (Ref.~\onlinecite{CadelanoE2009prl}) & $\sigma_{\rm int}=16.5$ Nm$^{-1}$ (Refs.~\onlinecite{CooperRC2013prb1,CooperRC2013prb2}) \tabularnewline
 & $\sigma_{\rm int}=44.4$ Nm$^{-1}$  (Ref.~\onlinecite{ReddyCD2006nano}) & $\sigma_{\rm int}=17.5$ Nm$^{-1}$ (Ref.~\onlinecite{TaoP2014jap}) \tabularnewline
 & $\sigma_{\rm int}^{\rm zig}=$ 40.5 Nm$^{-1}$, $\sigma_{\rm int}^{\rm arm}=$36.9 
Nm$^{-1}$ (Ref.~\onlinecite{LiuF2007prb}) & \tabularnewline
 & $\sigma_{\rm int}^{\rm zig}=36.9$ Nm$^{-1}$, $\sigma_{\rm int}^{\rm arm}=$30.2Nm$^{-1}$  (Ref.~\onlinecite{ZhaoH2009nl}) & \tabularnewline
\hline 
bending modulus & $D=1.17$ eV (Refs.~\onlinecite{OuyangZC1997,TuZC2002}), 1.4 eV (Refs.~\onlinecite{ArroyoM2002,LuQ2009}) & $D=9.61$ eV(Ref.~\onlinecite{JiangJW2013bend})\tabularnewline
\hline 
buckling strain & $\epsilon_{c}=-\frac{2.64}{L^{2}}$ & $\epsilon_{c}=-\frac{43.52}{L^{2}}$\tabularnewline
\hline 
\multirow{4}{*}{nanoresonator} & $f_{0}\times Q=6.3\times10^{11}$ Hz (10 K, Ref.~\onlinecite{ZandeAMVD}) & \tabularnewline
 & $f_{0}\times Q=1.82\times10^{12}$ Hz (5 K, Ref.~\onlinecite{ChenC2009nn}) & $f_{0}\times Q\approx2\times10^{9}$ Hz (300~K, Ref.~\onlinecite{Castellanos-GomezA2013adm})\tabularnewline
 & $f_{0}\times Q=1.56\times10^{13}$ Hz (90 mK, Ref.~\onlinecite{EichlerA}) & \tabularnewline
\cline{2-3} 
 & $f_{0}\times Q=6.4T^{-1.2}\times10^{3}$ THz (Ref.~\onlinecite{JiangJW2013mos2resonator}) & $f_{0}\times Q=2.4T^{-1.3}\times10^{4}$ THz (Ref.~\onlinecite{JiangJW2013mos2resonator}) \tabularnewline
\hline 
\multirow{5}{*}{thermal conductivity } & 5000 Wm$^{-1}$K$^{-1}$ (ballistic, $L=1\mu m$, Ref.~\onlinecite{JiangJW2009direction}) & 800 Wm$^{-1}$K$^{-1}$ (ballistic, $L=1\mu m$, Refs.~\onlinecite{HuangW,JiangJW2013mos2})\tabularnewline
\cline{2-3} 
 & 60 Wm$^{-1}$K$^{-1}$ ($L=10$ nm, Ref.~\onlinecite{JiangJW2010isotopic}) & 6 Wm$^{-1}$K$^{-1}$ ($L=4$ nm, Ref.~\onlinecite{JiangJW2013sw})\tabularnewline
 & 250 Wm$^{-1}$K$^{-1}$ ($L=300$ nm, Ref.~\onlinecite{XuX2014nc}) & 2 Wm$^{-1}$K$^{-1}$ ($L=120$ nm, Ref.~\onlinecite{LiuX2013apl})\tabularnewline
 & $\kappa>1500$ Wm$^{-1}$K$^{-1}$ ($L>4$ $\mu$m, Refs.~\onlinecite{BalandinAA2008,GhoshS2008apl,CaiWW2010nl,XuX2014nc}) & \tabularnewline
\cline{2-3} 
 & 1000 Wm$^{-1}$K$^{-1}$ (thick graphene layers, Ref.~\onlinecite{GhoshS}) & 52 Wm$^{-1}$K$^{-1}$ (thick MoS$_{2}$ layers, Ref.~\onlinecite{SahooS})\tabularnewline
\hline 
electronic band & Dirac cone; $E_{\rm gap}=0$ (Refs.~\onlinecite{NovoselovKS2005nat,ZhouSY2006np}) & parabolic; $E_{\rm gap}\approx1.8$ eV (direct, Refs.~\onlinecite{KamKK,EknapakulT2014nl})\tabularnewline
\hline 
\multirow{3}{*}{optical absorption} & fast photoresponse (Ref.~\onlinecite{XiaF2009nn}) & \tabularnewline
 & large band width (Ref.~\onlinecite{XiaF2009nn}) & \tabularnewline
 & low photoresponsivity (0.5 mAW$^{-1}$ Ref.~\onlinecite{XiaF2009nn}) & high photoresponsivity (880 AW$^{-1}$, Ref.~\onlinecite{Lopez-SanchezO2013nn})\tabularnewline
\hline 
\end{tabular}

\end{table*}

\section{Graphene/MoS$_{2}$ Heterostructure}
Thus far, we have focused on comparing several of the properties of graphene and SLMoS$_{2}$. The remainder of the article is devoted to studies on the close relationship between these two materials. As graphene and SLMoS$_{2}$ have complementary physical properties, it is natural to combine graphene and SLMoS$_{2}$ in specific ways to create heterostructures that mitigate the negative properties of each individual constituent.\cite{MaY2011ns,XiaF2012nns,BritnellL2013sci,RoyaK2013ssc,Algara-SillerG2013apl,ZanR2013acsn,MyoungN2013acsn,BertolazziS2013nl,ZhangW2013sr,LarentisS2014nl,ZhangW2014sr,XuH2014oe,WangLF2014nano}

There have been some experiments investigating the advanced properties of graphene/MoS$_{2}$ heterostructures. Britnell et al. found that graphene/MoS$_{2}$ heterostructures have high quality photon absorption and electron-hole creation properties, because of the enhanced light-matter interactions by the SLMoS$_{2}$ layer.\cite{BritnellL2013sci} Graphene has outstanding mechanical properties. These properties have been used to protect the MoS$_{2}$ films from radiation damage.\cite{ZanR2013acsn} Recently, Larentis et al. measured the electron transport in graphene/MoS$_{2}$ heterostructures and observed a negative compressibility in the MoS$_{2}$ component.\cite{LarentisS2014nl} This surprising phenomenon was interpreted based on the interplay between the Dirac and parabolic bands for graphene and MoS$_{2}$, respectively. Yu et al. fabricated high-performance electronic circuits based on the graphene/MoS$_{2}$ heterostructure, with MoS$_{2}$ as the transistor channel and graphene as the contact electrodes and the circuit interconnects.\cite{YuL2014nl}

Although experimentalists have shown great interest in graphene/MoS$_{2}$ heterostructures, the corresponding theoretical efforts have been limited until recently. First-principle calculations predicted that the inter-layer space and binding energy for the heterostructure are -21.0~{meV} and 3.66~{\AA} in Ref.~\onlinecite{MiwaRH2013jpcm} and -23.0~meV and 3.32~{\AA} in Ref.~\onlinecite{MaY2011ns}. Using these two quantities,\cite{JiangJW2014gmgyoung} a set of Lennard-Jones potential parameters were determined as $\epsilon$=3.95~{meV} and $\sigma$=3.625~{\AA}, with a cutoff of 10.0~{\AA}. These potential parameters were used to study the tension-induced structure transition of the graphene/MoS$_{2}$/graphene heterostructure. It was shown that the Young's modulus of the graphene/MoS$_{2}$/graphene heterostructure could be predicted by the following rule of mixtures, based on the arithmetic average,\cite{KarkkainenKK}
\begin{eqnarray}
Y_{\rm GMG} = Y_{G}f_{G} + Y_{M}f_{M},
\end{eqnarray}
where $Y_{\rm GMG}$, $Y_{G}$ and $Y_{M}$ are the Young's moduli of the heterostructure, graphene and SLMoS$_{2}$, respectively. $f_{G}=2V_G / (2V_G + V_M)=0.524$ is the volume fraction for the two outer graphene layers in the heterostructure and $f_{M}=V_M / (2V_G + V_M)=0.476$ is the volume fraction for the inner SLMoS$_{2}$ layer. The thicknesses were 3.35~{\AA} and 6.09~{\AA} for single-layer graphene and SLMoS$_{2}$, respectively. At room temperature, the Young's modulus was 859.69~GPa for graphene and 128.75~GPa for SLMoS$_{2}$. From this mixing rule, the upper-bound of the Young's modulus for the heterostructure was 511.76~GPa.

As another important mechanical property, the total strain of the graphene/MoS$_{2}$/graphene heterostructure was about 0.26, which is much smaller than the value of 0.40 for SLMoS$_{2}$.\cite{JiangJW2014gmgyoung} Under large mechanical tension, the heterostructure collapsed from the buckling of the outer graphene layers. These graphene layers were compressed in the lateral direction by the Poisson-induced stress, when the heterostructure was stretched in the longitudinal direction.

\section{Conclusions}
We have compared the mechanical properties of graphene and SLMoS$_{2}$. The main results are listed in Table~\ref{tab_summary}. This table serves as a resource for the prediction of corresponding properties for the graphene/MoS$_{2}$ heterostructure.

\textbf{Acknowledgements} This work was supported by the Recruitment Program of Global Youth Experts of China and start-up funding from Shanghai University.

%\bibliographystyle{aipnum4-1}
%\bibliographystyle{model3-num-names}
%\bibliographystyle{nature}
%\bibliography{/home/JiangJinWu/Documents/papers/mypapers/latex/biball}

\end{document}